\documentclass[aps,pra,twocolumn,superscriptaddress]{revtex4-1}

\usepackage{graphicx}
\usepackage{amsmath}

\begin{document}

\title{Fermionic ghost imaging}

\author{Jianbin Liu}
\email[]{liujianbin@xjtu.edu.cn}
\affiliation{Electronic Materials Research Laboratory, Key Laboratory of the Ministry of Education \& International Center for Dielectric Research, Xi'an Jiaotong University, Xi'an 710049, China}

\author{Yu Zhou}
\affiliation{MOE Key Laboratory for Nonequilibrium Synthesis and Modulation of Condensed Matter, Department of Applied Physics, Xi'an Jiaotong University, Xi'an 710049, China}

\author{Huaibin Zheng}
\affiliation{Electronic Materials Research Laboratory, Key Laboratory of the Ministry of Education \& International Center for Dielectric Research, Xi'an Jiaotong University, Xi'an 710049, China}

\author{Hui Chen}
\affiliation{Electronic Materials Research Laboratory, Key Laboratory of the Ministry of Education \& International Center for Dielectric Research, Xi'an Jiaotong University, Xi'an 710049, China}

\author{Fu-li Li}
\affiliation{MOE Key Laboratory for Nonequilibrium Synthesis and Modulation of Condensed Matter, Department of Applied Physics, Xi'an Jiaotong University, Xi'an 710049, China}

\author{Zhuo Xu}
\affiliation{Electronic Materials Research Laboratory, Key Laboratory of the Ministry of Education \& International Center for Dielectric Research, Xi'an Jiaotong University, Xi'an 710049, China}

\date{\today}

\begin{abstract}
Ghost imaging with thermal fermions is calculated based on two-particle interference in Feynman's path integral theory. It is found that ghost imaging with thermal fermions can be simulated by ghost imaging with thermal bosons and classical particles. Photons in pseudothermal light are employed to experimentally study fermionic ghost imaging. Ghost imaging with thermal bosons and fermions is discussed based on the point-to-point (spot) correlation between the object and image planes. The employed method offers an efficient guidance for future ghost imaging with real thermal fermions, which may also be generalized to study other second-order interference phenomena with fermions.
\end{abstract}

\maketitle

\section{Introduction}

Ghost imaging is a technique that obtains the image of an object by employing the second- or higher-order correlation of light. In a traditional ghost imaging scheme, a light beam is split into two and one beam is incident to the object, in which all the transmitted or reflected photons are collected by a bucket detector with no position information. The photons in the other beam, which does not interact with the object, is collected by a detector with position information. The image of the object can not be obtained by either signal from these two detectors alone. However, an image of the object can be retrieved if the signals from these two detectors are correlated. This peculiar property of the imaging technique is the reason why it is called ghost imaging \cite{shih-book}.

Ghost imaging was first realized with entangled photon pairs generated by spontaneous parametric down conversion from a nonlinear crystal \cite{klyshko,pittman}. It was first thought that entanglement is necessary for ghost imaging \cite{abouraddy}. However, inspired by Bennink \textit{et. al.}'s work \cite{bennink}, it was found that ghost imaging can also be realized with thermal light \cite{gatti-2004,chen-2004,valencia,cai}. Although there is no final conclusion about the physics of ghost imaging, the discussions on this very topic greatly improve the understanding about the second- and higher-order coherence of light \cite{shih-book,boyd}. Recent studies on ghost imaging are mainly focused on its possible applications. For instance, ghost imaging can be employed in both long range optical imaging \cite{zhao} and short range microscope \cite{yu,radwell}. High resolution ghost imaging \cite{gong} is possible when compressive sensing is employed \cite{katz}. Computational ghost imaging \cite{shapiro} shows great potential in single-pixel camera \cite{sun-2013,sun-2016,shin}. Ghost imaging with X-ray and atoms were also reported \cite{yu-2016,pelliccia,khakimov}. There are also many studies about other possible applications of ghost imaging \cite{sun-oe,chen-2013,li-oe,yao-oe,ryczkowski}. However, all the studies above are ghost imaging with bosons. There are limited number of studies that are ghost imaging with fermions \cite{gan,liu-2016}, both of which are theoretical investigations. To the best of our knowledge, there is no ghost imaging experiment with fermions due to the experiments with fermions are challenge \cite{henny,oliver}. Recently, it is suggested that the second-order interference of independent fermions can be simulated by the second-order interference of independent bosons and classical particles \cite{toppel,liu-arxiv}. In this paper, we will employ photons in pseudothermal light to experimentally study fermionic ghost imaging, which offers an efficient guidance for future ghost imaging with real thermal fermions.

This paper is organized as follows. In Sect. \ref{theory}, we will employ Feynman's path integral theory to calculate ghost imaging with thermal fermions. The experimental study of fermionic ghost imaging with photons is in Sect. \ref{experiment}. The discussions about ghost imaging with thermal bosons and thermal fermions are in Sect. \ref{discussion}. Section \ref{conclusion} summarizes our conclusions.

\section{Theory}\label{theory}

Two different theories can be employed to calculate ghost imaging with thermal light. One is classical intensity fluctuation correlation theory \cite{brown-1957,brown-1958}. The other one is quantum optical coherence theory mainly developed by Glauber \cite{glauber,glauber-1}. Both theories were originally developed to explain the famous two-photon bunching effect of thermal light discovered by Hanbury Brown and Twiss in 1956 \cite{HBT,HBT-1}. These two theories are equivalent for ghost imaging with classical light \cite{glauber,sudarshan}. However, only quantum theory is valid when nonclassical light is employed. Quantum theory is needed for ghost imaging with thermal fermions, since there is no fermions in classical theory.

There is another quantum theory to interpret the second-order interference of light besides Glauber's quantum optical coherence theory, which is two-photon interference in Feynman's path integral theory. This method was first employed by Fano to interpret the two-photon bunching effect of thermal light shortly after the effect was reported \cite{fano}. In fact, Feynman himself had also employed this method to interpret two-photon bunching effect in one of his lectures \cite{feynman-qed}. Recently, we have employed the same method to discuss the second-order interference of two independent light beams \cite{liu-pra,liu-oe,liu-epl,liu-cpb}, which greatly simplifies the calculation and offers a better understanding about the relation between the mathematical calculations and physical interpretations. We will also employ the same method to calculate ghost imaging with thermal fermions.

\begin{figure}[htb]
\centering
\includegraphics[width=50mm]{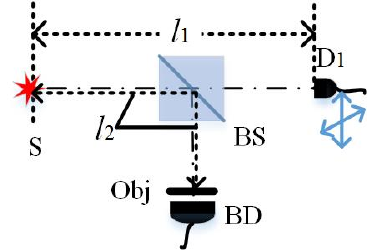}%
\caption{Scheme for ghost imaging with thermal fermions. S: particle sources. BS: 50:50 non-polarized beam splitter. Obj: object for imaging. BD: bucket detector. D$_1$: scannable detector. $l_1$: the distance between S and D$_1$. $l_2$: the optical distance between S and object via BS.}\label{1}
\end{figure}

The scheme for ghost imaging with thermal fermions is similar as the one for ghost imaging with thermal light \cite{valencia}, which is shown in Fig. \ref{1}. S is a thermal fermion source. BS is a 50:50 nonpolarized beam splitter. Obj is the object for imaging. BD is bucket detector, which collects all the fermions interacting with the object. D$_1$ is a two-dimension scannable detector. $l_1$ is the distance between the source and the scanning detector planes. $l_2$ is the distance between the source and object planes via BS. Since BD collects all the fermions interacting with object, we can first assume a point detector, D$_2$, is positioned in the object plane and then integrates over the object area to calculate the signal from BD.

There are two different alternatives for the particles emitted by thermal source S to trigger a two-particle coincidence count at D$_1$ and D$_2$ \cite{fano,feynman-qed,liu-pra,liu-oe,liu-epl,liu-cpb}, which is equivalent to a Hanbury Brown-Twiss (HBT) interferometer \cite{HBT,HBT-1}. One is particle a goes to D$_1$ and particle b goes to D$_2$. The other one is particle a goes to D$_2$ and particle b goes to D$_1$. If these two different alternatives are indistinguishable, the second-order coherence function of thermal fermions in a HBT interferometer is \cite{feynman}
\begin{eqnarray}\label{g2}
G^{(2)}_F(\mathbf{r}_1,t_1;\mathbf{r}_2,t_2)=\langle|A_{a1}A_{b2}-A_{a2}A_{b1}|^2\rangle,
\end{eqnarray}
where $(\mathbf{r}_j,t_j)$ is the space-time coordinates for the particle detection event at D$_j$ ($j=1$ and 2). $\langle...\rangle$ is ensemble average by taking all possible realizations into account. $A_{\alpha j}$ is the probability amplitude for particle $\alpha$ goes to D$_j$ ($\alpha=a$ and $b$, $j=1$ and 2). The minus sign in Eq. (\ref{g2}) is due to exchanging asymmetry of fermions \cite{feynman}. The image can be obtained by integrating the position of D$_2$ \cite{shih-book,scarcelli},
\begin{eqnarray}\label{g2-obj}
G^{(2)}_F(\vec{\rho}_1)=\int_{S_{Obj}}\langle|A_{a1}A_{b2}-A_{a2}A_{b1}|^2\rangle |T(\vec{\rho}_2)|^2d \vec{\rho}_2,
\end{eqnarray}
where $\vec{\rho}_j$ is transverse two-dimension position vector in D$_j$ plane ($j$=1 and 2). $S_{Obj}$ is the area of object and $T(\vec{\rho}_2)$ is the transmission function of the object.

As long as the particle is specified and the corresponding Feynman's particle propagator is substituted into Eq. (\ref{g2-obj}), fermionic ghost image of the object can be retrieved via $G^{(2)}(\vec{\rho}_1)$. In order to employ the well-known results of thermal light to simplify the calculations, we will assume there is a special type of fermion, which is called ``fermionic photon''. All the properties of ``fermionic photon'' are exactly the same as photon except ``fermionic photon'' has half integral spin and obey Fermi-Dirac statistics. The Feynman's propagator of ``fermionic photon'' is the same as the one of photon. Hence the results for the second-order interference of thermal light can be applied to the second-order interference of thermal ``fermionic photons'' by changing the plus sign into minus sign. Equation (\ref{g2}) can be simplified as  \cite{shih-book}
\begin{eqnarray}\label{g2-3}
g^{(2)}_F(\vec{\rho}_1-\vec{\rho}_2)= 1- \mbox{somb}^2(\frac{\pi d}{\lambda l}|\vec{\rho}_1-\vec{\rho}_2|),
\end{eqnarray}
where paraxial approximation and $l_1=l_2$ have been assumed to simplify the results. $g^{(2)}_F(\vec{\rho}_1-\vec{\rho}_2)$ is the normalized second-order coherence function \cite{glauber,glauber-1}. $\mbox{somb}(x)$ equals $2J_1(x)/x$, where $J_1(x)$ is the first-order Bessel function \cite{shih-book}. $d$ is the diameter of the source, which is assumed to be a circle in the calculation. $l$ is the distance between the source and object planes. $\lambda$ is wavelength of the employed ``fermionic photon''. In order to compare the experimental and theoretical results, the one-dimension case of Eq. (\ref{g2-3}) is \cite{shih-book}
\begin{eqnarray}\label{g2-3-1}
g^{(2)}_F(x_1-x_2) = 1- \mbox{sinc}^2[\frac{\pi d}{\lambda l}(x_1-x_2)],
\end{eqnarray}
where $\mbox{sinc}[\frac{\pi d}{\lambda l}(x_1-x_2)]$ equals $\mbox{sin}[\frac{\pi d}{\lambda l}(x_1-x_2)]/[\frac{\pi d}{\lambda l}(x_1-x_2)] $ and $x_1-x_2$ is the transverse relative distance between D$_1$ and D$_2$ in the detection planes. In this case, $d$ is the length of the one-dimension source.

With the help of Eq. (\ref{g2-3}), ghost imaging with thermal ``fermionic photons'' can be expressed as
\begin{eqnarray}\label{g2-4}
g^{(2)}_F(\vec{\rho}_1) = \int_{S_{Obj}} [1- \mbox{somb}^2(\frac{\pi d}{\lambda l}|\vec{\rho}_1-\vec{\rho}_2|)] |T(\vec{\rho}_2)|^2d \vec{\rho}_2.
\end{eqnarray}
When the size of thermal fermion source is large enough, $\mbox{somb}(\vec{\rho}_1-\vec{\rho}_2)$ can be approximated as $\delta(\vec{\rho}_1-\vec{\rho}_2)$. Substituting this approximation into Eq. (\ref{g2-4}), we can have
\begin{eqnarray}\label{g2-5}
g^{(2)}_F(\vec{\rho}_1) = c_1-|T(\vec{\rho}_1)|^2,
\end{eqnarray}
where $c_1$ is a constant background and equals $\int_{S_{Obj}}d \vec{\rho}_2$. The second term on the righthand side of Eq. (\ref{g2-5}) contains the information of the object, which is similar as the one of ghost imaging with thermal light. The only difference between ghost imaging with thermal light and thermal ``fermionic photons'' is the image is below the noise level in the latter.

In a recent work, T\"{o}ppel \textit{et. al.} proved that a pair of identical classical particles has exactly half fermionic and half bosonic characteristics, which means the two-particle interference of fermions can be simulated by two-particle interference of bosons and classical particles \cite{toppel}. Based on their work, we further proved that the second-order interference of thermal fermions in a HBT interferometer can be simulated by thermal bosons and classical particles in the same interferometer \cite{liu-arxiv}. Ghost imaging with thermal particles is based on the second-order interference of thermal particles in a HBT interferometer \cite{wang-2009}. The same conclusion as the one in Ref. \cite{toppel,liu-arxiv} should be true for ghost imaging with thermal particles. The detail calculations supporting the conclusion are in the following part.

The second-order coherence function of thermal light in a HBT interferometer is \cite{feynman,liu-arxiv}
\begin{eqnarray}\label{g2-b}
G^{(2)}_B(\mathbf{r}_1,t_1;\mathbf{r}_2,t_2)=\langle|A_{a1}A_{b2}+A_{a2}A_{b1}|^2\rangle,
\end{eqnarray}
where the meanings of all the symbols are similar as the ones in Eq. (\ref{g2}). The only difference between Eqs. (\ref{g2}) and (\ref{g2-b}) is the minus sign is changed into plus sign in the latter. With the same method above, it is straightforward to have the normalized second-order coherence function for thermal light ghost imaging in Fig. \ref{1} as
\begin{eqnarray}\label{g2-b1}
g^{(2)}_B(\vec{\rho}_1) = \int_{S_{Obj}} [1+ \mbox{somb}^2(\frac{\pi d}{\lambda l}|\vec{\rho}_1-\vec{\rho}_2|)] |T(\vec{\rho}_2)|^2d \vec{\rho}_2.
\end{eqnarray}

When classical particles are employed in a HBT interferometer, the second-order coherence function is
\begin{eqnarray}\label{g2-c}
G^{(2)}_C(\mathbf{r}_1,t_1;\mathbf{r}_2,t_2)=\langle|A_{a1}A_{b2}|^2+|A_{a2}A_{b1}|^2\rangle.
\end{eqnarray}
The reason why the probabilities instead of probability amplitudes are summed in Eq. (\ref{g2-c}) is due to the two different alternatives to trigger a two-particle coincidence count are distinguishable for classical particles. Assuming the propagator for classical particle is the same as the one of photon, the second-order coherence function for ghost imaging with classical particles in Fig. \ref{1} is
\begin{eqnarray}\label{g2-c1}
g^{(2)}_C(\vec{\rho}_1) = \int_{S_{Obj}} |T(\vec{\rho}_2)|^2d \vec{\rho}_2,
\end{eqnarray}
in which no image can be retrieved if classical particles were employed.

Comparing Eqs. (\ref{g2-4}), (\ref{g2-b1}), and (\ref{g2-c1}), it indeed that the relation,
\begin{eqnarray}\label{g2-6}
g^{(2)}_F(\vec{\rho}_1) =2 g^{(2)}_C(\vec{\rho}_1)-g^{(2)}_B(\vec{\rho}_1),
\end{eqnarray}
holds for thermal fermions, bosons and classical particles in Fig. \ref{1}, which is the same as the one in Refs. \cite{toppel,liu-arxiv}. Hence we can employ ghost imaging with thermal bosons and classical particles to simulate ghost imaging with thermal fermions.

\section{Experiments}\label{experiment}

\begin{figure}[htb]
\centering
\includegraphics[width=60mm]{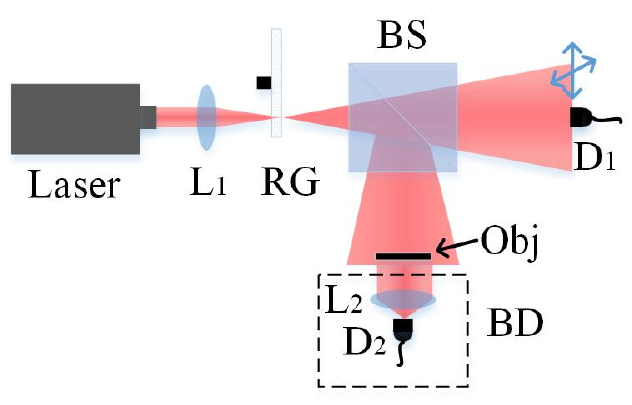}%
\caption{Experimental setup for ghost imaging with ``fermionic photons''. Laser: single-mode continuous-wave laser. L: lens. RG: Rotating ground glass. The meanings of other symbols are similar as the ones in Fig. \ref{1}. }\label{2}
\end{figure}

In this section, we will employ photons in pesudothermal light \cite{martienssen} to experimentally study ghost imaging with thermal ``fermionic photons''. The experimental setup is shown in Fig. \ref{2}, which is the same as ghost imaging with thermal light. Pesudothermal light is created by impinging a focused laser light beam onto a rotating ground glass (RG). The employed laser is a single-mode continuous-wave laser with central wavelength at 780 nm and frequency bandwidth of 200 kHz (Newport, SWL-7513). L$_1$ and L$_2$ are two lens with focus length of 50 mm. The meanings of other symbols are similar as the ones in Fig. \ref{1}. D$_1$ and D$_2$ are two single-photon detectors (PerkinElmer, SPCM-AQRH-14-FC).  BD consists of a single-photon detector and a focus lens, L$_2$. The distance between the S and D$_1$ planes equals the one between S and object planes, which is 910 mm.

\begin{figure}[htb]
\centering
\includegraphics[width=60mm]{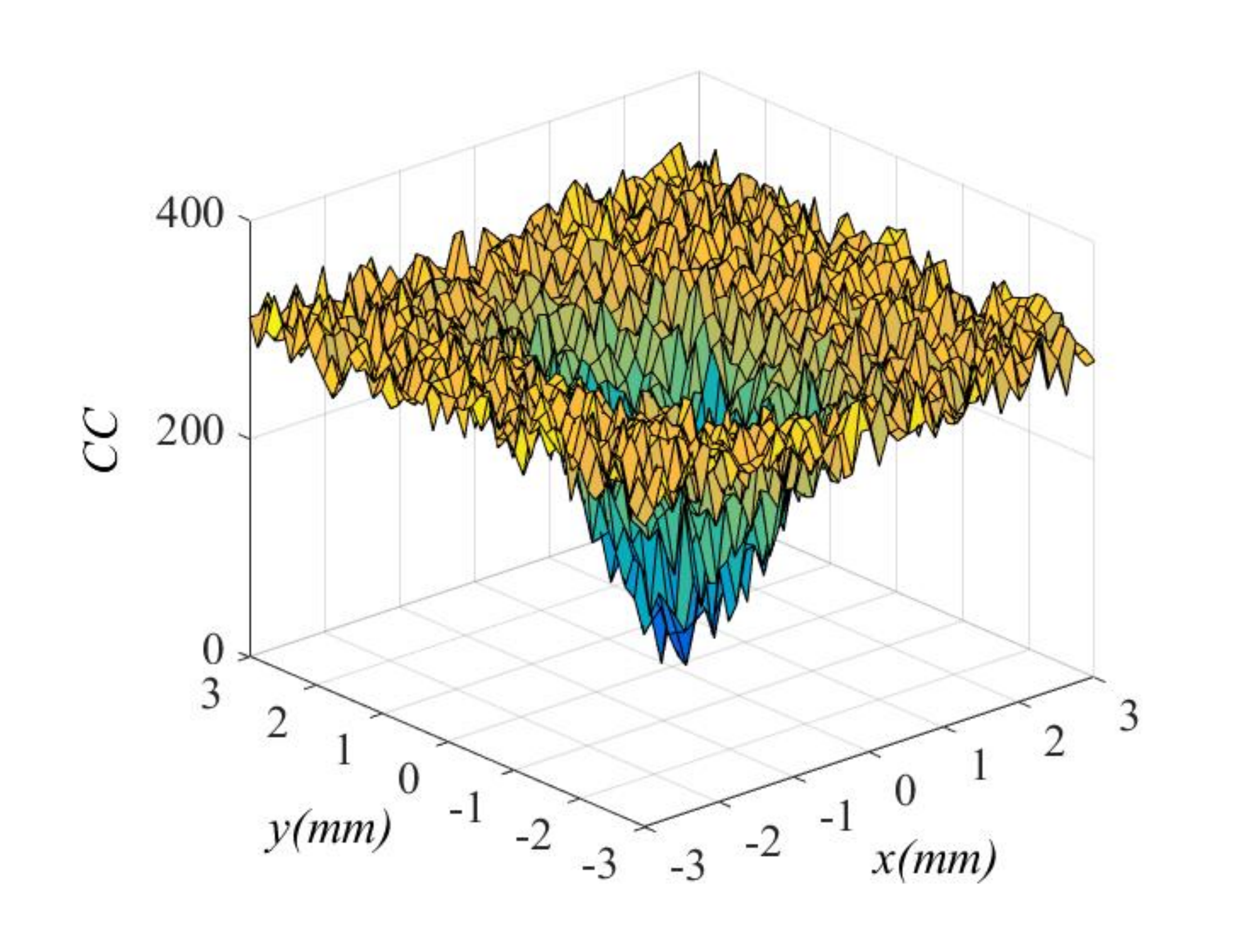}%
\caption{Calculated two-particle coincidence counts of thermal ``fermionic photons'' in a HBT interferometer. $x$ and $y$ are two transverse spatial coordinates of the scanning detector, respectively. $CC$ is two-particle coincidence counts. }\label{3}
\end{figure}

We first measure the second-order spatial correlation function of thermal ``fermionic photons'' by moving D$_2$ to the object plane and transversely scanning the position of D$_1$, which is equivalent to a HBT interferometer. Two single-mode fibers with diameter of 5 $\mu$m are connected two single-photon detectors in the measurement, respectively. The measuring time for each position is 30 s, which is much longer than the second-order coherence time of pseudothermal light.  The scanning step is 0.125 mm. The two-particle coincidence counts for thermal ``fermionic photons'' in a HBT interferometer are shown in Fig. \ref{3}, which is proportional to $g^{(2)}_F(\vec{\rho}_1-\vec{\rho}_2)$ and calculated by employing Eq. (\ref{g2-6}). $g^{(2)}_B(\vec{\rho}_1-\vec{\rho}_2)$ is proportional to the measured two-photon coincidence counts of pseudothermal light. $g^{(2)}_C(\vec{\rho}_1-\vec{\rho}_2)$ is proportional to the constant background coincidence counts in the same measurement.  When these two detectors are at symmetrical positions, the two-particle coincidence count gets its minimum as shown in Fig. \ref{3}. Different from two-photon bunching in thermal light \cite{HBT,HBT-1}, two-particle antibunching is observed for thermal ``fermionic photons''. The calculated two-particle antibunching effect is similar as the measured results with thermal electrons in a HBT interferometer \cite{henny,oliver}. The full width of half maximum (FWHM) of the dip determines the resolution of fermionic ghost imaging, which is similar as the one in thermal light ghost imaging.

\begin{figure}[htb]
\centering
\includegraphics[width=90mm]{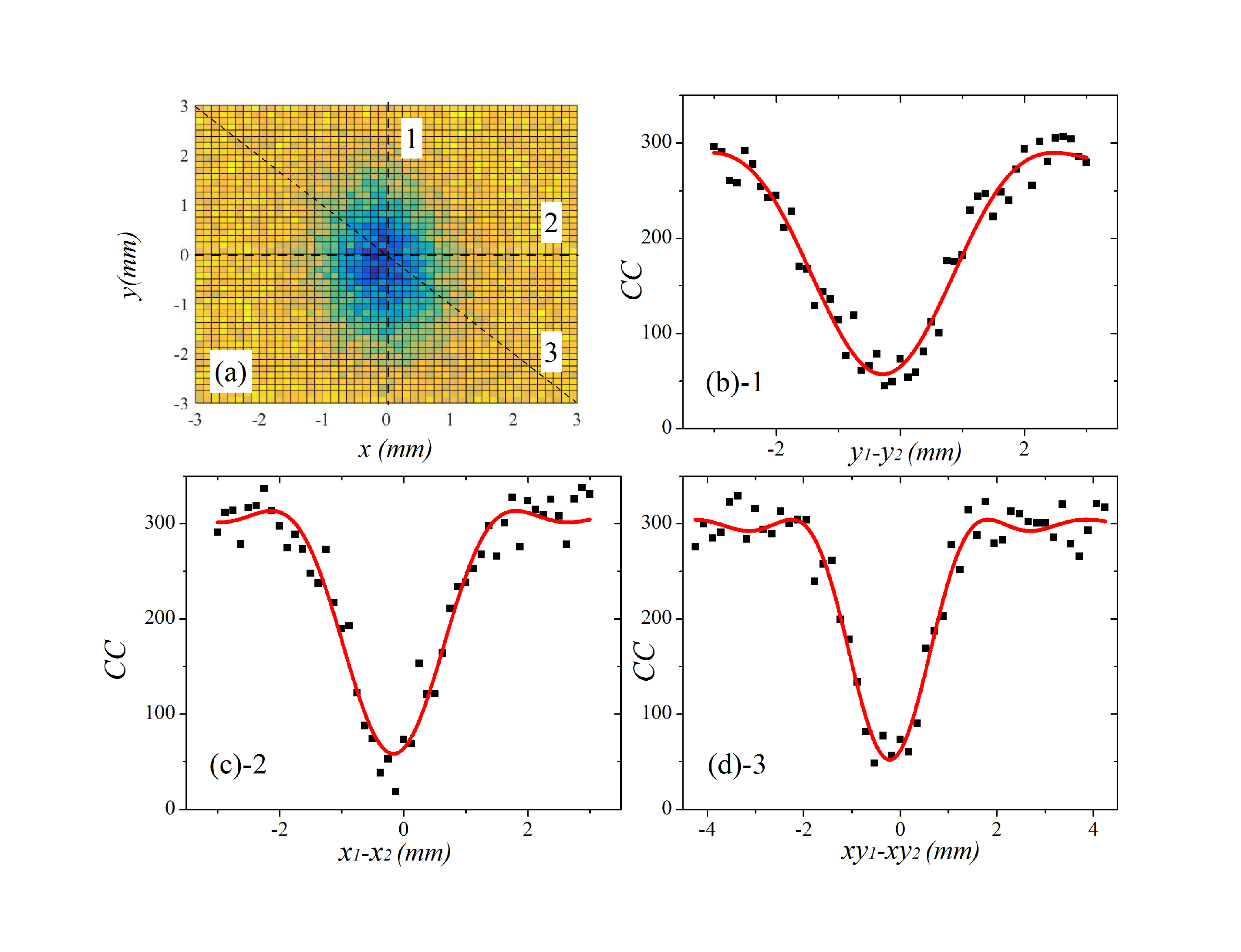}%
\caption{Detail analysis of the second-order coherence function of thermal ``fermionic photons'' in a HBT interferometer. (a) is the top view of the second-order coherence function in Fig. \ref{3}. (b), (c) and (d) are the one-dimension second-order coherence functions along lines 1, 2 and 3 in (a), respectively. $y_1-y_2$ is the relative distance between D$_1$ and D$_2$ in $y$-axis. $x_1-x_2$ is the relative distance between D$_1$ and D$_2$ in $x$-axis. $xy_1-xy_2$ is the relative distance between D$_1$ and D$_2$ along line 3. The black squares are the calculated two-particle coincidence counts. The red lines are theoretical fittings of the data by employing Eq. (\ref{g2-3-1}).}\label{4}
\end{figure}

Figure \ref{4}(a) is the top view of the calculated two-particle coincidence counts in Fig. \ref{3}. The reason why the measured antibunching dip is not circular symmetric is that the pseudothermal light source in our experiment is not circular symmetric. In order to analyze the details about the second-order coherence function of thermal ``fermionic photons'', we will discuss three different situations as shown in Figs. \ref{4}(b), (c), and (d), respectively. The black squares are two-particle coincidence counts and the red curves are theoretical fitting by employing Eq. (\ref{g2-3-1}). Figure \ref{4}(b), (c) and (d) are the one-dimension second-order coherence functions along lines 1, 2 and 3 in Fig. \ref{4}(a), respectively. The FWHM of the fitted curves in Figs. \ref{4} (b), (c) and (d) are $0.77 \pm 0.05$, $0.55 \pm 0.03$ and $0.57 \pm 0.03$ mm, respectively. The lengths of light source in these three directions are $0.26 \pm 0.02$, $0.36 \pm 0.02$ and $0.35 \pm 0.02$ mm, respectively. It is obvious that higher resolution in one direction corresponds to larger size of light source in the corresponding direction, which is similar as the conclusion in thermal light ghost imaging \cite{shih-book}.

\begin{figure}[htb]
\centering
\includegraphics[width=80mm]{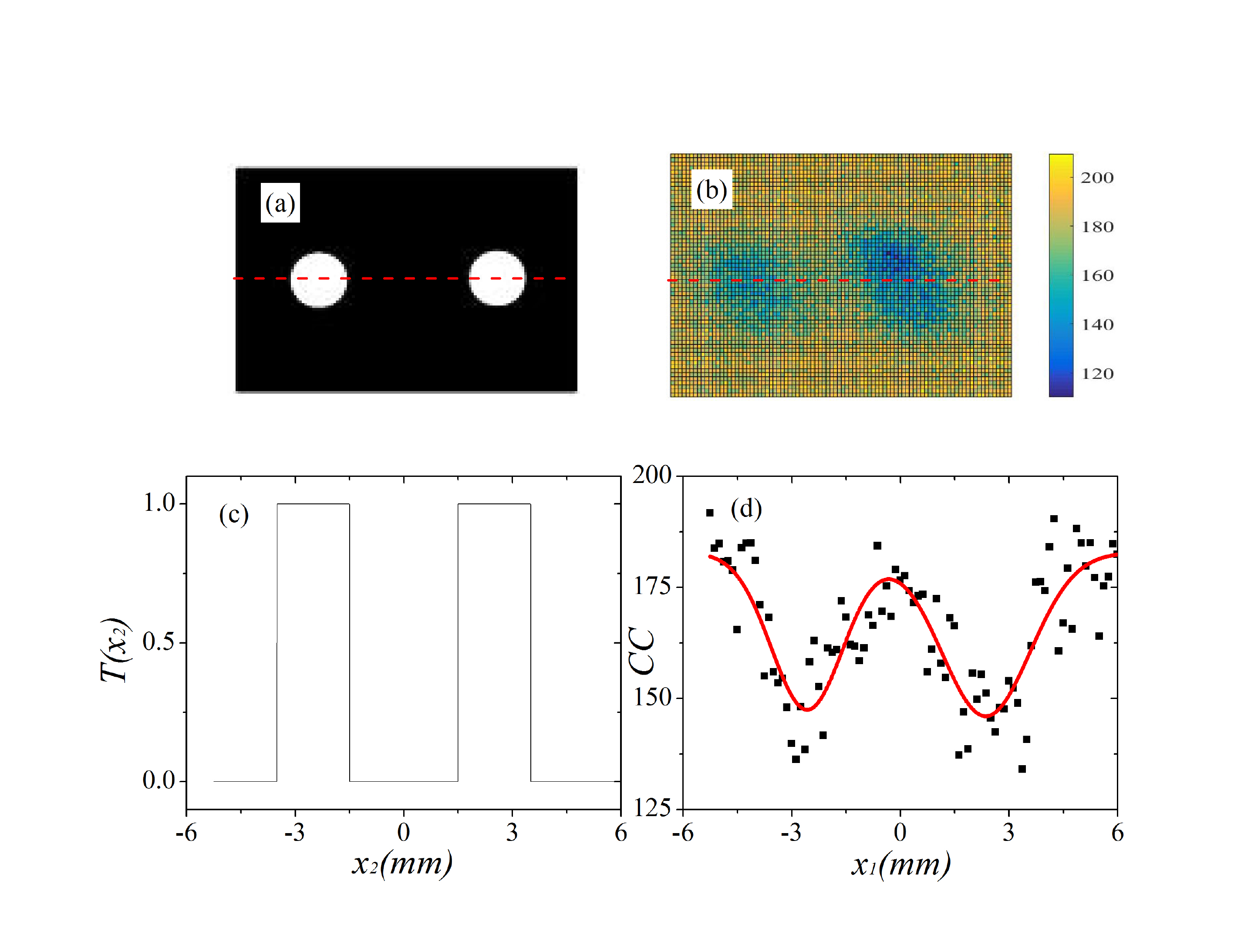}%
\caption{Fermionic ghost imaging. (a) and (c) are objects for imaging. (b) and (d) are re-constructed fermionic ghost images for (a) and (c), respectively. (c) and (d) are the one-dimension functions along the red dash lines in (a) and (b), respectively. The reasons why the calculated ghost images deviating from the original images are due to the resolution of our system is about the same size of the object for imaging and the spatial anticorrelation function is not circular symmetry.}\label{5}
\end{figure}

Figure \ref{5} is the result of ghost imaging with thermal ``fermionic photons''. The bucket detector consists of a single-photon detector connected to a multimode fiber with diameter 200 $\mu$m and a focus lens. The object is transmissive double pinholes as shown in Fig. \ref{5}(a). The diameters of both pinholes are 2 mm and the distance between the center of these two pinholes is 5 mm. The re-constructed image of the object is shown in Fig. \ref{5}(b), which is calculated in the same method as the one in Fig. \ref{3}. Unlike ghost imaging with thermal light that the signals of the object is above the noise level \cite{valencia,scarcelli,boyd,shapiro}, the signals of fermionic ghost imaging is below the noise level. Figure \ref{5}(c) is the one-dimension object along the red line in Fig. \ref{5}(a). In Fig. \ref{5}(d), the black squares are the calculated two-particle coincidence counts for ``fermionic photons'' and the red line are theoretical fitting by employing Gaussian functions. The transmitting parts in Fig. \ref{5}(c) of the object correspond to two dips in Fig. \ref{5}(d) of the re-constructed images. The reasons why the re-constructed images deviating from the object are as follows. The first reason is that the measured second-order spatial coherence function in Fig. \ref{3} is not circular symmetric, which will cause the re-constructed image not be strict circles as the object. The second reason is that the resolutions of the imaging system are at the same level of the object's size so that the border of the object can not be clearly reflected. However, the distance between the centers of two dips in Fig. \ref{5}(d) equals $5.0 \pm 0.3$ mm, which is consistent the distance between the centers of two pinholes of the object.

\section{Discussions}\label{discussion}

\begin{figure}[htb]
\centering
\includegraphics[width=70mm]{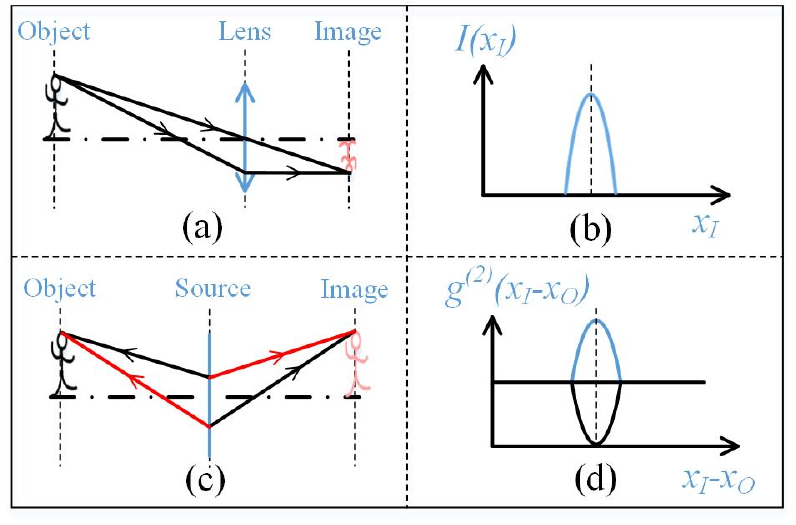}%
\caption{Imaging principle. (a) classical imaging with focus lens. (b) point-to-spot correlation of intensity in the object and image planes in (a). $x_I$ is the coordinate of image plane and $I(x_I)$ is the intensity distribution in the image plane. (c) ghost imaging with thermal particles. The two red lines and two black lines are two different alternatives for two particles emitted by thermal source to trigger a two-particle coincidence count, respectively. (d) point-to-spot correlation of the second-order coherence function for thermal bosons (peak) and fermions (dip). $x_I-x_O$ is transverse position difference between the coordinates in the image and object planes, respectively. $g^{(2)}(x_I-x_O)$ is the normalized second-order coherence function. }\label{6}
\end{figure}

In order to understand why the image of the object in fermionic ghost imaging is below the noise level, we will discuss how the image of an object is usually formed in a imaging system. In a traditional imaging system with lens as shown in Fig. \ref{6}(a), all the light fields emitted or reflected from one point in the object will interferes constructively at one and only one point in the image plane after passing though the lens. The image is perfect if the lens is perfect and infinity large, since there is point-to-point correlation between the object and image planes, which means a point in the object will form a point in the image. However, lens has finite size and the point-to-point correlation is changed into a point-to-spot correlation as shown in Fig. \ref{6}(b). The FWHM of the spot is usually treated as the resolution of the imaging system, which is determined by the lens, the distance between the object and lens, the wavelength of light and so on.

Figure \ref{6}(c) indicates there are two possible alternatives for two particles emitted by thermal source to trigger a two-particle coincidence count. Only when the distance between the source and object planes equals the one between the source and the image planes, a point-to-spot correlation can be formed via two-particle interference \cite{shih-book}. Two-photon bunching is observed for thermal light and the signals of the image in ghost imaging with thermal light is above the noise level. On the other hand, two-particle antibunching is observed for thermal fermions and the signal of fermionic ghost image should be below the noise level.

The point-to-spot correlation does not only originate from interference, but also originates from projection in classical optics. The point-to-spot correlation via projection can also be employed to form ghost image of an object by analogy of the usual ghost imaging scheme. For instance, the very early ``two-photon'' coincidence image by Bennink \textit{et. al.} is based on the point-to-spot correlation via projection \cite{bennink}. Recent computational ghost imaging experiments with projector or DMD are also based on the point-to-spot correlation via projection \cite{sun-2013,sun-2016,yu,radwell}. Even though both the projection ghost imaging and usual ghost imaging can form the image of the object and the experimental setups look similar, there is difference between these two types of ghost imaging. The difference is exactly the same as the one between imaging with lens and projection in classical optics. In the usual ghost imaging scheme and classical imaging with lens, the point-to-spot correlation is formed via interference. In the projection ghost imaging scheme and projection, the point-to-spot correlation is due to light travels in straight line.

In fact, no matter how the correlation is formed, an image can be formed as long as there is a point-to-point (spot) correlation between two planes. Recent reported quantum imaging with undetected photons can be understood in this way \cite{lemos}. Their result seems strange at first time. How can the image of the object be formed when both the detected photons did not interact with the object? In a related experiment, Zou \textit{et. al.} proved that the second-order interference effect can be observed only when it is impossible to tell which nonlinear crystal emits the detected photons \cite{zou}. There exists a point-to-point correlation if we put an object between these two crystals as the one in Ref. \cite{lemos}. When the corresponding point is blocked, there is no second-order interference effect. When there is no object in the corresponding point, there is second-order interference effect. This point-to-point correlation between whether there is an object and whether there is the second-order interference effect is the reason why they can form the image of the object in their experimental setup \cite{lemos}.

\section{Conclusions}\label{conclusion}

In summary, we have theoretically proved that fermionic ghost imaging can be simulated by ghost imaging with thermal bosons and classical particles. The second-order spatial coherence function of thermal ``fermionic photons'' is calculated based on the measurements for photons, in which a dip instead of a peak is observed. Fermionic ghost imaging of double pinholes is observed, where the signal is below the noise level. As long as there is point-to-point (spot) correlation between the object and image planes, the scheme can be employed to form the image of the object.  Employing thermal bosons and classical particles to study the behavior of thermal fermions offers an efficient guidance for future ghost imaging with real thermal fermions, which is possible in nowaday technology. The second-order spatial coherence function of thermal fermions have been measured in a HBT interferometer \cite{jeltes}, which is the foundation for fermionic ghost imaging and paves the way for ghost imaging with real fermions.

\section*{Acknowledgments}
This project is supported by National Science Foundation of China (No.11404255), Doctoral Fund of Ministry of Education of China (No.20130201120013), the 111 Project of China (No.B14040) and the Fundamental Research Funds for the Central Universities.

\end{document}